\documentclass[11pt]{article}
\usepackage{amsfonts}
\usepackage{amssymb}
\usepackage{mathrsfs}

\title{\rightline{\normalsize DO-TH 11/29} \bigskip
\bf Lepton number violating effects in neutrino oscillations}

\author{
{\bf Sebastian Hollenberg} \\ 
Fakult\"at f\"ur Physik, Technische Universit\"at Dortmund,\\ 
D-44221 Dortmund, Germany\\
\and \\
{\bf Octavian Micu} \\
Institute for Space Sciences,
P.O. Box MG–23, \\
Ro 077125 Bucharest-Magurele, Romania
\and \\
{\bf Palash B. Pal} \\
Saha Institute of Nuclear Physics, Kolkata, India
}

\date{}

\evensidemargin=\oddsidemargin
\def\bg#1{\mathchoice
{{\mbox{\boldmath $#1$}}}
{{\mbox{\boldmath $#1$}}}
{{\mbox{\boldmath $\scriptstyle #1$}}}
{\mbox{\boldmath $\scriptscriptstyle #1$}}} 

\def\bra#1{\left< #1 \right|}
\def\ket#1{\left| #1 \right>}
\def\inprod#1#2{\left< #1 | #2 \right>}
\def\tr{\mathop{\rm tr}}
\def\tilde{\widetilde}
\def\anu{\bar\nu}
\def\Eqn#1{Eq.~(\ref{#1})}
\def\Eqs#1#2{Eqs.~(\ref{#1}) and (\ref{#2})}

\font\euler=cmmi8 scaled 1300
\def\eu#1{\mbox{\euler #1}}
\def\aeu#1{\hat{\mbox{\euler #1}}}
\font\bbbold=msbm10 scaled 1100
\def\bb#1{\mbox{\bbbold #1}}
\def\ms#1{\mathscr #1}

\begin{document}

\maketitle

\begin{abstract}
  We develop a {\em non-adiabatic} perturbation theory for oscillations
  involving an arbitrary number of neutrino and antineutrino species,
  including the possibility of lepton number violation which we treat
  as a small effect.  We interpret the physics of such an approach for
  the one generation case by introducing the notion of adiabaticity
  for neutrino and antineutrino oscillations in analogy to flavor
  oscillations.  We find that in a CP-odd matter environment a small
  lepton number violation in vacuo can be enhanced.  Eventually, we
  apply the perturbation theory to the two generation case and work
  out an example for manifestations of lepton number violation, which
  can be solved both perturbatively as well as analytically thereby
  further clarifying the nature of the perturbation expansion.
\end{abstract}

Over the last decade or so the physics of neutrino flavor oscillations
has seen considerable developments.  Experiments have found that the
three neutrinos discovered so far undergo flavor oscillations.  Two of
the three associated flavor mixing angles have been measured, the
third one has been constrained to be small, but potentially non-zero.
Moreover, the two mass splittings between the neutrino mass
eigenstates have been determined.  Despite all progress, a few
neutrino properties remain elusive.  It is, for instance, yet unknown
whether there is any CP violation in the lepton sector or whether
neutrino mass eigenstates obey a normal or inverted hierarchy
\cite{Amsler:2008zzb}.

There are, however, also some experimental findings in neutrino
oscillation experiments which cannot be reconciled with the standard
description of neutrino flavor oscillations.  The latter findings are
therefore often referred to as {\em neutrino oscillation anomalies}.
Such findings as reported by the LSND \cite{Aguilar:2001ty} and
MiniBooNE \cite{AguilarArevalo:2007it, AguilarArevalo:2010wv}
collaborations have given rise to a plethora of theoretical
speculations how to explain them.  The latter range from additional
sterile neutrino species, non-standard neutrino interactions
\cite{Nelson:2010hz} and extra spatial dimensions \cite{Pas:2005rb} to
CPT violation in the lepton sector \cite{Kostelecky:2003cr}.

Neutrino oscillations depend on neutrino masses.  One of the most
important questions connected with neutrino masses is the possibility
of lepton number violation and consequent Majorana nature of the
neutrinos.  Lepton number violation should therefore affect neutrino
oscillations, and in fact the effects can come in two forms.  First of
all, in addition to flavor oscillations, one can have lepton number
violating oscillations, i.e., processes which can be termed
neutrino-antineutrino oscillations \cite{Wu:1980px}.  Second, the
usual flavor oscillation probabilities might be modified by the
presence of lepton number violating terms in the Lagrangian.

Despite the fact that the reported neutrino oscillation anomalies
might not last and may eventually be refuted by future neutrino
oscillation experiments, the question of whether physics beyond the
established picture of neutrino flavor oscillations with similar
signatures exists is ultimately an experimental one.  In this letter
we propose a generic parameterization for lepton number violating
neutrino flavor oscillations in order to qualitatively understand
their ramifications and identify deviations from the signals predicted
by lepton number conserving oscillations.

Let us, for such purposes, start by writing down the most general form
of the Schr\"odinger equation involving active neutrino and
antineutrino fields in flavor space.  We denote the left-chiral,
active neutrino fields collectively by the boldfaced symbol $\bg \nu
= (\nu_e, \nu_\mu, \nu_\tau, \dots)$.  The active antineutrino fields
are right-chiral, and they are denoted by $\bg \anu = (\anu_e,
\anu_\mu, \anu_\tau, \dots)$.  The Schr\"odinger equation has the form
\begin{eqnarray}
{\mathrm d \over \mathrm dt} \pmatrix {\bg\nu(t) \cr \bg\anu(t) } =
-i \bb H(t) \pmatrix {\bg\nu(t) \cr \bg\anu(t) } \,,
\label{SchEq}
\end{eqnarray}
where $\bb H$ is the Hamiltonian.  We assume that the energy of the
neutrino beams are such that the sterile states, which are expected to
be heavy if a see-saw mechanism is assumed to be responsible for
neutrino mass generation, are not excited and thus dynamically
decouple from neutrino flavor oscillations.

In the block form that has been used to write this equation, we can
write the total Hamiltonian in the form
\begin{eqnarray}
\bb H(t) = \pmatrix{ \eu H(t) & 0 \cr 0 & \aeu H(t)} +
\pmatrix{0 & \eu B(t) \cr \eu B^\dagger(t) & 0} \equiv \bb H_0(t) +
\delta \bb H(t) \,,
\label{H}
\end{eqnarray}
where we have separated out the block-diagonal and off-diagonal parts.
Since the fields in $\bg\nu$ are assigned lepton number equal to $+1$,
whereas those in $\bg\anu$ have the opposite lepton number, it implies
that $\bb H_0$ is the lepton number conserving part of the
Hamiltonian, whereas $\delta \bb H$ is lepton number violating.  While
writing the explicit form for $\delta \bb H$, we have used the
assumption that all neutrinos are stable particles, so that the
Hamiltonian is Hermitian.  This assumption also implies that the block
matrices $\eu H$ and $\aeu H$ are both Hermitian.  Note, moreover,
that $\eu H$ is generically different from $\aeu H$, since neutrino
and antineutrino fields can be subject to CP non-conserving
interactions as is the case, for instance, in elastic forward
scattering of neutrinos off background
fields~\cite{Wolfenstein:1977ue}.

If CPT is a symmetry of the Hamiltonian, there are additional
constraints on the elements of $\bb H$.  These constraints can be
identified by recognizing that the active antineutrino states are CPT
transforms of the active neutrino states.  In general, if an operator
$\ms O(t)$ commutes with CPT denoted by the operator $\Theta$, we can
write
\begin{eqnarray}
\bra{\Theta a} \ms O(t) \ket{\Theta b} = \bra b \ms O^\dagger(t) \ket a =
{\bra a \ms O(t) \ket b}^* 
\label{CPT}
\end{eqnarray}
for arbitrary state vectors $\ket a$ and $\ket b$, and denoting the
state $\Theta \ket b$ by $\ket{\Theta b}$.  For the active neutral
fermions which appear in \Eqn{SchEq}, we can choose the ordering as
well as the phases of the states in such a way that
\begin{eqnarray}
\Theta \ket{\nu_a} = \ket{\anu_a} \,.
\end{eqnarray}
Therefore, \Eqn{CPT}, along with the Hermiticity of the Hamiltonian,
would then imply the relation
\begin{eqnarray}
\bra{\nu_a} \bb H(t) \ket{\nu_b} =  \bra{\Theta\nu_b} \bb H(t)
\ket{\Theta\nu_a} = \bra{\anu_b} \bb H(t) \ket{\anu_a} \,.
\end{eqnarray}
In terms of the block-diagonal notation introduced in \Eqn{H}, this
can be written as
\begin{eqnarray}
\aeu H(t) = \eu H^\top(t) \,.
\label{HHhat}
\end{eqnarray}
Going through similar steps, we can also show that
\begin{eqnarray}
\eu B(t) = \eu B^\top(t) \,,
\label{Bsym}
\end{eqnarray}
i.e., the block $\eu B$ is symmetric.  Of course lepton number
violating neutrino-antineutrino oscillations can occur even in
scenarios with CPT violation, in which case \Eqs{HHhat}{Bsym} need not
hold; but also for the case in which \Eqs{HHhat}{Bsym} hold, i.e., CPT
is an exact symmetry, still lepton number violating neutrino
oscillations are possible.

For the purposes that concern us in this paper, it is more convenient
to introduce the time evolution operator $\bb U(t,t_0)$ via 
\begin{eqnarray}
\pmatrix {\bg\nu(t) \cr \bg\anu(t) } = \bb U(t,t_0) 
\pmatrix {\bg\nu(t_0) \cr \bg\anu(t_0) } \, .
\end{eqnarray}
The time evolution equation thus takes the form
\begin{eqnarray}
 \frac{\mathrm d}{\mathrm d t} \bb U(t,t_0) = -i \bb H(t) \bb U(t,t_0).
\end{eqnarray}
Henceforth, we will often take $t_0=0$ and will omit it in the
notation.  Let us have a look at the CPT properties of the time
evolution operator.  CPT symmetry would imply that the evolution
operator satisfies the condition
\begin{eqnarray}
\Theta \bb U(t) = \bb U(-t) \Theta \,,
\end{eqnarray}
and therefore the analog of \Eqn{CPT} would be
\begin{eqnarray}
\bra{\Theta a} \bb U(t) \ket{\Theta b} = {\bra a \bb U(-t) \ket b}^*
\,.  
\label{CPTU}
\end{eqnarray}
In particular, it says that
\begin{eqnarray}
\inprod {\anu_b} {\anu_a(t)} = \inprod {\nu_b} {\nu_a(-t)}^* = \inprod
{\nu_a} {\nu_b(t)} \,, 
\end{eqnarray}
implying that the oscillation probability of an $\anu_a$ going to
$\anu_b$ is the same as that of a $\nu_b$ going to $\nu_a$:
\begin{eqnarray}
P (\anu_a \to \anu_b; t) = P (\nu_b \to \nu_a; t) \,.
\end{eqnarray}
In addition, \Eqn{CPTU} also implies that 
\begin{eqnarray}
P (\nu_a \to \anu_b; t) = P (\nu_b \to \anu_a; t) \,.
\end{eqnarray}

In our analysis, we adopt the paradigm that lepton number violating
neutrino-antineutrino couplings $\eu B$ are small compared to lepton
number conserving terms $\eu H$, $\aeu H$.  We shall, in fact, adopt
$\eu B$ as a {\em small\/} quantity for a perturbative solution
\cite{Diaz:2009qk} for the time evolution.  To this end, we decouple
the evolution of neutrinos and antineutrinos from the evolution of the
neutrino-antineutrino system by introducing a (subsidiary) time
evolution operator $\bb G(t)$ for the neutrino and antineutrino
systems.  The latter is then used to transform the time evolution
equation for $\bb U(t,t_0)$ to what we shall henceforth refer to as
the {\em interaction picture}.  Establishing this line of action, we
write
\begin{eqnarray}
\bb U(t,t_0) = \bb G(t,t_0)\, \bb U_{\rm I}(t,t_0) \,,
\label{UI}
\end{eqnarray}
with
\begin{eqnarray}
\frac{\mathrm d}{\mathrm d t} \bb G(t,t_0) = -i \bb H_0(t) \bb G(t,t_0),
\label{dGdt}
\end{eqnarray}
where the subscript $\rm I$ indicates the associated quantity in
the interaction picture.  It is readily seen that the time evolution
equation in the interaction picture is given by
\begin{eqnarray}
 \frac{\mathrm d}{\mathrm d t} \bb U_{\rm I}(t,t_0) 
= -i \bb H_{\rm I} (t) \bb U_{\rm I}(t,t_0) \,,
\end{eqnarray}
with
\begin{eqnarray}
\bb H_{\rm I}(t) = \bb G^{-1}(t,t_0) \delta \bb H(t) \bb G(t,t_0).
\label{HI}
\end{eqnarray}
Since the Hamiltonian $\bb H_0$ is lepton number conserving, the
solution of \Eqn{dGdt} is obviously of the form
\begin{eqnarray}
\bb G(t,t_0) &=& \pmatrix { \eu G(t,t_0) & 0 \cr 0 & \aeu G(t,t_0)  }
\,.  
\end{eqnarray}
Putting this solution into \Eqn{HI}, we obtain
\begin{eqnarray}
 \bb H_{\rm I}(t) &=& \pmatrix {  0 & \eu G^{-1}(t,t_0) \eu B(t) 
\aeu G(t,t_0)
 \cr \aeu G^{-1}(t,t_0) \eu B^\dagger(t) \eu G(t,t_0) & 0 
  } \,.
\label{HIblock}
\end{eqnarray}

We can now perturbatively solve the time evolution equation in the
interaction picture by means of the Magnus expansion
\cite{Magnus:1954}.  We write its solution as a matrix exponential
\begin{eqnarray}
\bb U_{\rm I}(t,t_0) = e^{\Omega(t,t_0)} \,,
\label{Magnus}
\end{eqnarray}
where the Magnus operator $\Omega(t,t_0)$ is the sum of the so-called
Magnus approximants $\Omega_i(t,t_0)$ according to
\begin{eqnarray}
\Omega(t,t_0) \equiv \Omega_1(t,t_0) + \Omega_2(t,t_0) + \dots \, ,
\end{eqnarray}
which are {\em small} in an appropriate sense as we shall see in due
course.  Various methods have been worked out to calculate the Magnus
approximants \cite{Klarsfeld:1989}.  They are found to obey
\begin{eqnarray}
 \Omega_1(t,t_0) &=& -i \int_{t_0}^{t} \mathrm d \tau \; 
\bb H_{\rm I}(\tau), 
\label{Omega1}\\ 
 \Omega_2(t,t_0) &=& -\frac12 \int_{t_0}^{t} \mathrm d t_1
 \int_{t_0}^{t_1} \mathrm dt_2 \;
 \Big[\bb H_{\rm I}(t_1), \bb H_{\rm I}(t_2)\Big],
\label{Omega2}
\end{eqnarray}
up to second order in the perturbation.  We shall see in our analysis
that neutrino-antineutrino oscillation effects only enter the
oscillation probabilities in second order perturbation theory.

Our aim is to comment on how lepton number violating effects manifest
themselves.  We assume that the lepton number conserving effects,
encoded in $\bb G$, can be obtained either exactly or perturbatively
by solving the underlying Schr\"odinger equation.  The solution for
the time evolution operator $\bb U(t,t_0)$ can then be written as
\begin{eqnarray}
\bb U(t,t_0) = \bb G(t,t_0) e^{\Omega(t,t_0)} \,,
\end{eqnarray}
according to \Eqs{UI}{Magnus}.  Inserting the block form of the
interaction picture Hamiltonian from \Eqn{HIblock} into
\Eqs{Omega1}{Omega2}, we find that the Magnus operator up to second
order perturbation theory can be written as
\begin{eqnarray}
 \Omega^{(2)}(t,t_0) = \pmatrix{ \tilde {\eu C}(t,t_0) & -i\tilde {\eu
     B}(t,t_0) \cr 
           -i\tilde {\eu B}^\dagger(t,t_0) & \tilde {\eu D}(t,t_0) },
\end{eqnarray}
where we have introduced the shorthand notations
\begin{eqnarray}
 \tilde {\eu B}(t,t_0) &\equiv& \int_{t_0}^{t} \mathrm d \tau \; 
\eu G^{-1}(\tau,\tau_0) \eu B(\tau) \aeu G(\tau,\tau_0), 
 \label{Btilde} \\
 \tilde {\eu C}(t,t_0) &\equiv& -\frac{1}{2} \int_{t_0}^{t} \mathrm d t_1
 \int_{t_0}^{t_1} \mathrm d t_2 \;  
 \left\{\dot{\tilde {\eu B}}(t_1) \dot{\tilde {\eu B}}^\dagger(t_2) 
 - \left(\dot{\tilde {\eu B}}(t_1) \dot{\tilde {\eu B}}^\dagger(t_2)\right)^\dagger\right\}
                        \,, 
 \label{Ctilde} \\
 \tilde {\eu D}(t,t_0) &\equiv& -\frac{1}{2} \int_{t_0}^{t} \mathrm d t_1
 \int_{t_0}^{t_1} \mathrm d t_2 \; 
 \left\{\dot{\tilde {\eu B}}^\dagger(t_1) \dot{\tilde {\eu B}}(t_2) 
 - \left(\dot{\tilde {\eu B}}^\dagger(t_1) \dot{\tilde
     {\eu B}}(t_2)\right)^\dagger\right\} 
 \label{Dtilde} \,.
\end{eqnarray}
The dot indicates a derivative with respect to time.  Note that the
quantities $\tilde {\eu C}(t,t_0)$ and $\tilde {\eu D}(t,t_0)$ are
quadratic in $\tilde {\eu B}(t,t_0)$, i.e., quadratic in the lepton
number violating parameter $\eu B$ in the Hamiltonian.  Then, up to
second order terms in $\eu B$, we obtain
\begin{eqnarray}
e^{\Omega(t,t_0)} &\simeq& 1 + \Omega + \frac12 \Omega^2 +
\mathcal{O}(\Omega^3) \\
&=& \pmatrix{
1 + \mathcal{O}^{(2)}(t,t_0) & -i\tilde {\eu B}(t,t_0) \cr
                 -i\tilde {\eu B}^{\dagger}(t,t_0) & 1 + 
                 \hat{\mathcal O}^{(2)}(t,t_0)  }  + \cdots \,, 
\end{eqnarray}
where
\begin{eqnarray}
\mathcal O^{(2)}(t,t_0) &=& \tilde {\eu C}(t,t_0) - \frac{1}{2} \tilde {\eu B}(t,t_0)
\tilde {\eu B}^\dagger(t,t_0) \,, \label{O1}\\ 
\hat{\mathcal O}^{(2)}(t,t_0) &=& \tilde {\eu D}(t,t_0) - \frac{1}{2} \tilde
{\eu B}^\dagger(t,t_0) \tilde {\eu B}(t,t_0) \,.
\label{O2}
\end{eqnarray}
These quantities show how lepton number violating
neutrino-antineutrino mixing encoded in $\eu B(t)$ affects the
neutrino and antineutrino sectors respectively.

With these ingredients, we can now write down the oscillation
probabilities of various kinds.  Let us denote the oscillation
probabilities in absence of lepton number violating terms by the
notations
\begin{eqnarray}
P_0 (\nu_a \to \nu_b; t) &=& \Big| {\eu G}_{ba}(t) \Big|^2 \,, \nonumber\\ 
P_0 (\anu_a \to \anu_b; t) &=& \Big| {\aeu G}_{ba}(t) \Big|^2 \,.
\end{eqnarray}
Up to second order in lepton number violating terms, the probabilities
would then read
\begin{eqnarray}
P(\nu_a \to \nu_b; t) &=& P_0 (\nu_a \to \nu_b; t) + 2
\textrm{Re}\left\{ \eu G^{\ast}_{ba}(t) \times
       \Big( \eu G(t) \mathcal O^{(2)}(t) \Big)_{ba}
     \right\} , 
                         \label{nunu} \\  
P(\anu_a \to \anu_b; t) &=& P_0 (\anu_a \to \anu_b; t) + 2
\textrm{Re}\left\{ \aeu G^{\ast}_{ba}(t) \times 
       \Big( \aeu G(t) \hat{\mathcal O}^{(2)}(t) \Big)_{ba}
     \right\} \,.
\label{nubnub}
\end{eqnarray}
Note that there is no summation over recurring indices implied here
or elsewhere in the paper. Thus the lepton number preserving
oscillation probabilities get modified.  On the other hand, there can
also be lepton number violating oscillations now, and the
probabilities of such oscillations are given by
\begin{eqnarray}
P(\nu_a \to \anu_b; t) &=&  \Big| \Big(\aeu G(t) \tilde{\eu
  B}^\dagger(t)\Big)_{ba} \Big|^2  \,, 
\label{nunub}\\ 
P(\anu_a \to \nu_b; t) &=& \Big| \Big(\eu G(t) \tilde{\eu
  B}(t)\Big)_{ba} \Big|^2 .  \label{nubnu}
\end{eqnarray}
These oscillation probabilities for lepton number preserving as well
as violating neutrino (antineutrino) oscillations Eqs.\
(\ref{nunu}--\ref{nubnu}) present the main result of our analysis.  A
few comments are in order.

First, we want to emphasize that this method for obtaining the
oscillation probabilities does not involve the effective mixing
matrix at any stage of the analysis.  It is of course possible, in
principle at least, to diagonalize the entire Hamiltonian including
the lepton number violating terms and derive oscillation probabilities
from the resulting eigenvalues and eigenstates.  Our aim was to
demonstrate how some of the essential features of such a solution can
be obtained without performing such a diagonalization.  Note also that
this approach reduces a $2{\eu N} \times 2 {\eu N}$ problem (${\eu N}$
being the number of neutrino species involved), the diagonalization of
the Hamiltonian, to a ${\eu N} \times {\eu N}$ problem, since the
perturbation expansion only involves block entries from the original
Hamiltonian $\bb H(t)$.

Truncating the exponential of the Magnus operator results in loss of
unitarity.  Since we truncate the series in a way that we only keep
terms up to second order in the small perturbation $\eu B(t)$, we also
expect that unitarity of the time evolution operator is only conserved
up to second order in this quantity.  It is straightforward to verify
this by invoking the unitarity condition on $\bb U(t,t_0)$.  Note that
such considerations also pertain to solutions for the neutrino and
antineutrino evolution operators $\eu G(t,t_0)$ and $\aeu G(t,t_0)$,
if their solution is obtained in a similar way, i.e., by truncating
the exponential of the associated Magnus operator in a perturbative
manner.  The fact that we seek a perturbative expansion of the
solution to the time evolution operator readily implies that the
structure of the oscillation probabilities is of the same kind.
However, it should be realized that the probabilities given here are
viable approximations for times that satisfy $|\eu B|t \ll 1$, where
$|\eu B|$ indicates the magnitude of any non-zero eigenvalue of $\eu
B$.  Hence the loss of unitarity is small as it is triggered through
the small quantity $\tilde {\eu B}(t)$, and will appear only at orders
${\eu B}^3$ or higher which have been neglected in writing down the
expression for various oscillation probabilities.

In fact, unitarity can be tested explicitly from the expressions of
the oscillation probabilities given above. Take, for example,
\Eqn{nunu}.  It shows that
\begin{eqnarray}
\sum_b P(\nu_a \to \nu_b; t) &=& 
1 + 2 \textrm{Re} \sum_b \left\{ \eu G^{\ast}_{ba}(t) \times
       \Big( \eu G(t) \mathcal O^{(2)}(t) \Big)_{ba}
     \right\} \nonumber\\
&=& 1 + 2 \textrm{Re} \Big( \eu G^\dagger(t) \eu G(t) \mathcal O^{(2)}(t)
\Big)_{aa} = 1 + 2 \textrm{Re} \Big( \mathcal O^{(2)}(t)
\Big)_{aa} \,,
\end{eqnarray}
using the unitarity of the matrix $\eu G$.  Further, looking at the
expression of $\mathcal O^{(2)}(t)$ in \Eqn{O1}, we see that the
diagonal elements of this matrix must have negative real part, since
the diagonal elements of $\tilde {\eu C}$ are purely imaginary and
those of $\tilde {\eu B} \tilde {\eu B}^\dagger$ must be real and
positive. Thus, starting from a particular flavor of neutrino $\nu_a$,
the total probability for lepton number conserving oscillations is
\begin{eqnarray}
\sum_b P(\nu_a \to \nu_b; t) 
&=& 1 - \Big( \tilde {\eu B}(t)
\tilde {\eu B}^\dagger(t) \Big)_{aa} \,,
\end{eqnarray}
which is less than unity, as one should expect.  It should also be
noted, from \Eqn{nunub}, that with the same initial state, the total
probability for lepton number violating oscillations is given by
\begin{eqnarray}
\sum_b P(\nu_a \to \anu_b; t) &=& \sum_b \Big( \aeu G(t) \tilde{\eu
  B}^\dagger(t) \Big)^*_{ba} \Big( \aeu G(t) \tilde{\eu
  B}^\dagger(t) \Big)_{ba} = \Big( \tilde{\eu
  B}(t) \aeu G^\dagger(t) \aeu G(t) 
\tilde{\eu B}^\dagger (t)\Big)_{aa} \nonumber\\
&=& \Big( \tilde{\eu
  B}(t) \tilde{\eu B}^\dagger(t)\Big)_{aa} \,,
\end{eqnarray}
using the unitarity of the matrix $\aeu G$.  Therefore
\begin{eqnarray}
\sum_b P(\nu_a \to \nu_b; t) + \sum_b P(\nu_a \to
\anu_b; t) = 1 \,,
\end{eqnarray}
as expected, implying that in presence of lepton number violation, the
lepton number conserving oscillation probabilities should decrease,
making room for lepton number violating oscillations.

The form of the oscillation probabilities given in Eqs.\
(\ref{nunu}--\ref{nubnu}) is illustrative in the sense that for the
limiting case of vanishing neutrino-antineutrino coupling $\eu B(t)
\to 0$ it readily reduces to the standard, lepton number conserving
neutrino oscillation results.  The appearance of the oscillation
probabilities for $\nu_a \to \anu_b$ and $\anu_a \to \nu_b$ can also
be interpreted in an intuitive way: reading Eqs. (\ref{nunub} --
\ref{nubnu}) from right to left they state that $\tilde {\eu
  B}^\dagger(t)$ ($\tilde {\eu B}(t)$) switches the initial neutrino
(antineutrino) state to the associated antiparticle and $\aeu G (t)$
($\eu G (t)$) then evolves the antiparticle (particle) state until its
detection.  So neutrino-antineutrino oscillations are clearly a signal
for lepton number violation in the neutrino sector.  Note that this
statement is a general one: all oscillation probabilities come with
$\tilde {\eu B}(t)$ and derivatives thereof; put another way, in an
approach in which neutrino-antineutrino couplings are treated as a
small perturbation to all other potential enhancements in the neutrino
and antineutrino sector respectively, one cannot have modifications in
the neutrino-neutrino and antineutrino-antineutrino probabilities
without introducing neutrino-antineutrino conversions at the same
time.  The oscillation probabilities for $\nu_a \to \anu_b$ and
$\anu_a \to \nu_b$ are generically different by virtue of the generic
difference between $\aeu G(t)$ and $\eu G(t)$.  Even if the respective
Hamiltonians $\eu H$ and $\aeu H$ do not discriminate between
particles and antiparticles (e.g.  $\eu H(t) = \aeu H(t)$) there still
is a difference due to the fact that neutrino-antineutrino coupling
can be complex and not self-adjoint, i.e.  $\tilde {\eu B}(t) \neq
\tilde {\eu B}^{\dagger}(t)$ in general.

A similar situation pertains to $\nu_a \to \nu_b$ and $\anu_a \to
\anu_b$ oscillations.  They differ in the time evolution operators for
neutrinos and antineutrinos, but also in the second order operators
$\mathcal{O}^{(2)}(t)$ and $\hat{\mathcal{O}}^{(2)}(t)$.  Those
operators are, in general, not identical, which is again due to the
fact that $\tilde {\eu B}(t)$ does not have to be self-adjoint.  This
statement translates to the fact that lepton number violating neutrino
oscillations discriminate between particles and antiparticles even if
there is no difference in the respective Hamiltonians for neutrinos
and antineutrinos.  This behavior is also expected from lepton number
violation; note, that this distinguishing feature is generated
dynamically rather than being imposed by hand.

Having established an approach to neutrino oscillations in which
lepton number violation is considered a {\em small} effect, let us now
elaborate on the physical notion underlying this framework.  For such
purposes we begin by studying the one generation case of neutrino
oscillations, since already in this case lepton number violation
allows neutrino-antineutrino oscillations to develop.  Making use of
the formalism, we obtain
\begin{eqnarray}
 \tilde {\eu B}(t,t_0) = \int_{t_0}^{t} \mathrm d \tau ~ {\eu B}(\tau)
 \, e^{i\Delta\tilde {\eu H}(\tau)} , 
\label{B1gen}
\end{eqnarray}
where we have defined 
\begin{eqnarray}
 \Delta\tilde {\eu H}(t) \equiv 
 \int_{t_0}^{t} \mathrm d \tau \; \Delta H(\tau) =
 \int_{t_0}^{t} \mathrm d \tau
 \left[{\eu H}(\tau) - {\aeu H}(\tau)\right]  
\end{eqnarray}
as the difference between (CP non-conserving) {\em potential} terms
in the neutrino and antineutrino sectors.  Clearly, if the difference
in such potential terms vanishes for some time $t = t_{\rm res}$, the
integral \Eqn{B1gen} has a stationary phase and we can evaluate it by
means of the saddle point approximation.  This yields
\begin{eqnarray}
 \tilde {\eu B} \simeq e^{i\Delta\tilde {\eu H}(t_{\rm res})} \sqrt{\frac{\pi}{2}\frac{1}{\gamma_{\rm res}}}.
\label{B1gensol}
\end{eqnarray}
Here we have introduced the {\em adiabaticity parameter at resonance}
$\gamma_{\rm res}$, which we shall properly define and explain
shortly.  To this end, let us start from the Hamiltonian of \Eqn{H}
and notice that for a simple two dimensional case, we can always
easily find a unitary transformation (for a two dimensional problem
phases are irrelevant and the unitary transformation amounts to a 
time-dependent rotation in flavor space), which entails an {\em effective
  mixing angle} $\Theta(t)$ fixed via
\begin{eqnarray}
 \cos\Theta(t) = \frac{\Delta {\eu H}(t)}{\omega_{\rm eff}}, \qquad
 \sin\Theta(t) = \frac{2\vert {\eu B}(t)\vert}{\omega_{\rm eff}}.
\end{eqnarray}
The {\em effective oscillation frequency} $\omega_{\rm eff}$ of the
system is here given by
\begin{eqnarray}
 \omega_{\rm eff} = \sqrt{4\vert{\eu B}(t)\vert^2 + \Delta{\eu H}^2(t)}.
\end{eqnarray}
It is interesting to note that the effective mixing becomes maximal,
neutrino conversions undergo a resonance, in the case of vanishing
$\Delta {\eu H}(t)$, i.e.  $\Delta {\eu H}(t_{\rm res}) = 0$ at some
resonance time $t_{\rm res}$.  Put another way, mixing between
neutrinos and antineutrinos becomes maximal when the difference
between potential terms for the two species is minimal (or in fact
vanishing at the resonance time).  This is also corroborated by the
fact that for identical neutrino and antineutrino potentials mixing is
always maximal.  Obviously, for the case of CP conserving neutrino and
antineutrino potentials the difference between those vanishes at all
times since they are identical to begin with.  However, in the case of
CP non-conserving matter potentials, the difference is generally
non-zero and can be varying with time also --- the latter effect is
crucial for some physics implications of the resonance structure as
shall be seen in due course.  It is also conceivable to allow for CPT
violating potential terms, which then generate a non-vanishing
difference $\Delta H(t)$.

With the effective mixing at our disposal, we write the adiabaticity
parameter at resonance as
\begin{eqnarray}
 \gamma(t_{\rm res}) \equiv \gamma_{\rm res} = \frac{1}{\omega_{\rm eff}} 
 \left.\frac{\mathrm d \Theta(t)}{\mathrm d t}\right|_{t=t_{\rm res}} =
 - \frac{1}{4\vert 
{\eu B}(t) \vert^2} 
 \left.\frac{\mathrm d \Delta {\eu H}(t)}{\mathrm d t}\right|_{t=t_{\rm res}}.
\label{gamma}
\end{eqnarray}

This definition of an adiabaticity parameter can also be understood
physically as follows: the characteristic time of the
neutrino-antineutrino system is $\tau_{\rm sys} \sim 1/\omega_{\rm
  eff}$, whereas the characteristic time of the interaction can be
given as $\tau_{\rm int} \sim (\mathrm d \Theta / \mathrm d t)^{-1}$.
Hence a small $\gamma_{\rm res}$ states that the system's time scale
is much smaller than the interaction's time scale.  Put another way,
the transition is adiabatic.

For $\gamma_{\rm res} \gg 1$ we encounter non-adiabatic
neutrino-antineutrino conversions; $\gamma_{\rm res} \ll 1$ gives the
adiabatic case.  So roughly speaking (neglecting the derivative of
the potential term for the time being) the smaller ${\eu B}(t)$, the
larger the non-adiabaticity of the neutrino-antineutrino system and
hence if we start the evolution with only $\nu_a$ ($\anu_a$) states
present the transition $\nu_a \to \nu_a$ ($\anu_a \to \anu_a$)
prevails; the lepton number violating oscillation channel $\nu_a \to
\anu_a$ ($\anu_a \to \nu_a$) gets more and more suppressed as the
non-adiabaticity increases.  If, however, the change of the difference
in matter potentials is small with time, this effect can partially
compensate a small neutrino-antineutrino coupling, driving the
evolution of the system towards adiabatic transitions opening the
lepton number violating oscillation channel $\nu_a \to \anu_a$
($\anu_a \to \nu_a$) again.  Note also that the suppression of the
lepton number violating oscillation channel depends on the time
dependence of the potential terms in the Hamiltonian.

Substituting the result for the adiabaticity parameter $\gamma_{\rm
  res}$ of \Eqn{gamma} in the expression for $\tilde{\eu B}$ of
\Eqn{B1gensol} and at the same time keeping in mind that the
oscillation probability $P(\nu_a \to \anu_a; t)$ of \Eqn{nunub}
is directly proportional to $\tilde{\eu B}^\dagger$, it is seen that
interpreting lepton number violating (neutrino-antineutrino) couplings
as a small perturbation to lepton number conserving neutrino
oscillations in the one generation case is equivalent to assuming that
neutrino-antineutrino oscillations occur non-adiabatically;
non-adiabaticity hence {\em closes} the lepton number violating
oscillation channel $\nu_a \to \anu_a$, but at the same time improves
the perturbative expansion as outlined above.

These results for the one generation framework make the case for
referring to the perturbation theory as developed in this letter as a
{\em non-adiabatic perturbation expansion}.

From the definition of the adiabaticity parameter $\gamma_{\rm res}$
in \Eqn{gamma} another interesting feature emerges.  Suppose we assume
{\em small} lepton number violating couplings ${\eu B}(t)$ between
neutrinos and antineutrinos in the oscillation Hamiltonian.  This
means that oscillations between particles and antiparticles are
suppressed by the large non-adiabaticity of the transitions giving
rise mostly to the lepton number conserving oscillation channel $\nu_a
\to \nu_a$.  If, however, the time variation of the difference in
matter potentials at the resonance is sufficiently mild, it is seen
that the presence of such matter along the neutrino (antineutrino)
propagation path can drive the system towards adiabaticity thus
opening the oscillation channel $\nu_a \to \anu_a$.  Put another way,
the presence of CP (or even CPT) non-conserving matter of a varying
density (readily re-interpreted as a time dependence of the potential
terms) can enhance lepton number violating neutrino oscillations as
compared to the case in vacuo.  Note that this statement holds
regardless of the nature of the perturbation theory developed in this
letter, since it is merely a result obtained from the adiabaticity
parameter $\gamma_{\rm res}$ and the latter does only need the
Hamiltonian of the system as a prerequisite.

Let us next approach the two generation case in vacuo.  In this case,
apart from a term proportional to the unit matrix, the flavor
oscillation of neutrinos as well as antineutrinos is governed by the
Hamiltonian
\begin{eqnarray}
\eu H = \aeu H = {\Delta m^2 \over 4p} \Big( \sigma_x \sin 2\theta  -
\sigma_z \cos 2\theta \Big) \equiv \omega \Big( \sigma_x \sin 2\theta  -
\sigma_z \cos 2\theta  \Big) \,, 
\end{eqnarray}
where the $\sigma$'s are the usual Pauli matrices, and $\theta$ is the
mixing angle in absence of lepton number violating terms.
Accordingly, the evolution operator in the neutrino as well as the
antineutrino sector will be given by
\begin{eqnarray}
\eu G(t,t_0=0) = \aeu G(t,t_0=0) = \cos \omega t - i
\Big( \sigma_x \sin 2\theta  -
\sigma_z  \cos 2\theta \Big)
\sin \omega t \, ,
\end{eqnarray}
in absence of lepton number violating terms.  In order to include the
effects of the lepton number violating terms, let us first write the
matrix ${\eu B}(t)$ as
\begin{eqnarray}
{\eu B} (t) =  b_0(t) + b_i(t) \sigma_i  \,,
\label{b0bi}
\end{eqnarray}
where 
\begin{eqnarray}
b_0(t) = \frac12 \tr {\eu B}(t) \,, \qquad b_i(t) = 
\frac12 \tr \Big( \sigma_i {\eu
  B}(t) \Big) \,.
\end{eqnarray}
Note that if CPT is conserved, \Eqn{Bsym} dictates that we must have
$b_y(t)=0$.  It is straightforward to see that 
\begin{eqnarray}
\dot{\tilde {\eu B}}(t) = \beta_0(t) + \beta_i(t) \sigma_i \,,
\end{eqnarray}
where
\begin{eqnarray}
\beta_0(t) &=& b_0(t) \,, \nonumber\\ 
\beta_x(t) &=& \sin 2\theta \Big[ b_x(t) \sin2\theta - b_z(t) \cos2\theta
\Big] + \cos 2\omega t \cos 2\theta \Big[  b_x(t) \cos2\theta + b_z(t) 
\sin2\theta \Big] \nonumber\\*  && 
- b_y(t) \sin 2 \omega t \cos 2\theta  \,, \nonumber\\  
\beta_y(t) &=& b_y(t) \cos 2\omega t + \sin 2\omega t \Big[ b_x(t) \cos 2\theta + b_z(t) \sin 2\theta
\Big] \,,  \nonumber \\ 
\beta_z(t) &=& - \cos 2\theta \Big[ b_x(t) \sin2\theta - b_z(t) \cos2\theta
\Big] + \cos 2\omega t \sin 2\theta \Big[  b_x(t) \cos2\theta + b_z(t) 
\sin2\theta \Big] \nonumber\\* && - b_y(t) \sin 2\omega t \sin 2\theta  \,.
\label{betas}
\end{eqnarray}
We will assume here that the elements of $\eu B$ are time-independent.
These expressions can then be easily integrated to give $\tilde {\eu 
  B}$, and the resulting expressions plugged into the relevant
formulas to obtain various oscillation probabilities.

For the most general $\eu B$, the expressions are cumbrous.  Here, we
present an illustrative example for a very simple form of $\eu B$
which can be analytically tackled without much trouble.  Consider $\eu
B$ to be the multiple of a unit matrix, which means that the lepton
number violating effects are somehow flavor blind.  In \Eqn{b0bi},
this would mean that all $b_i=0$, only $b_0$ is non-zero.  Now, from
\Eqn{betas}, we find that $\tilde{\eu B}$ is $b_0t$ times the unit
matrix.  Looking back at \Eqs{nunub}{nubnu}, we find that the
probabilities of neutrino-antineutrino oscillations are given by
\begin{eqnarray}
P(\nu_a \to \anu_b; t) = |b_0|^2 t^2 \times P_0 (\nu_a \to
\nu_b; t) \,, 
\label{bsoln1}
\end{eqnarray}
whereas the lepton number conserving oscillation probabilities are
given by
\begin{eqnarray}
P(\nu_a \to \nu_b; t) = \Big( 1 - |b_0|^2 t^2 \Big) \times P_0 (\nu_a
\to \nu_b; t) \,.
\label{bsoln2}
\end{eqnarray}

Let us discuss the nature of this solution.  If there is no
lepton number violation, a plot of oscillation probability vs.\ $t$
would exhibit periodic highs and lows of the same magnitude.  If there
is lepton number violation, these periodic nature would be modulated
by the function $(1 - |b_0|^2 t^2)$.  This means that, as the path
length increases, the successive maxima and minima of probability
would be less and less pronounced.  There is, of course, no fear of
the maxima and minima vanishing and the curve becoming completely
flat, because, as said earlier, these solutions should be valid in the
region where $|b_0| t \ll 1$.

In fact, with this particular choice of the matrix $\eu B$, one can
do better and solve the problem exactly.  To see this, let us recall
the form of the interaction Hamiltonian given in \Eqn{HI}.  Since $\eu
G = \aeu G$ for the Hamiltonian $\bb H_0$, we obtain the $2\times2$
block structure of $\bb H_{\rm I}$ as follows:
\begin{eqnarray}
\bb H_{\rm I} = \pmatrix {  0 & b_0 
 \cr b_0^* & 0 } = |b_0| \pmatrix {  0 & e^{i\alpha} 
 \cr e^{-i\alpha} & 0 } \,,
\end{eqnarray}
where $b_0 = |b_0|e^{i\alpha}$.  Moreover, in the exponent in the
Magnus expansion, only the term $\Omega_1$ is non-zero.  From
\Eqn{Omega1}, we find that it is given by
\begin{eqnarray}
\Omega(t) = \Omega_1(t) = -i\bb H_{\rm I} t \,.
\end{eqnarray}
Because of the special form for the matrix $\bb H_{\rm I}$, the
exponentiation can be done exactly, and one obtains
\begin{eqnarray}
e^{\Omega(t)} = \pmatrix{1 & 0 \cr 0 & 1}  \cos |b_0|t - i \pmatrix {
  0 & e^{i\alpha} 
 \cr e^{-i\alpha} & 0 } \sin |b_0|t \,.
\end{eqnarray}

The evolution operator of the system is defined through 
\Eqs{UI}{Magnus}.  It shows that
\begin{eqnarray}
P(\nu_a \to \nu_b; t) &=& \cos^2 |b_0|t \times P_0 (\nu_a
\to \nu_b; t) \,, \nonumber\\*
P(\nu_a \to \anu_b; t) &=& \sin^2 |b_0|t \times P_0 (\nu_a
\to \nu_b; t) \,.
\label{exact}
\end{eqnarray}
The expressions given in \Eqs{bsoln1}{bsoln2} are nothing but the
approximations up to leading order terms in the time dependence.
However, in the form given here in \Eqn{exact}, it is valid for all
$t$. Note that, although this last illustration uses a
time-independent Hamiltonian, the general formulas that we give in
Eqs.\ (\ref{nunu}--\ref{nubnu}) are valid for time-dependent cases as
well, e.g., in matter induced oscillations involving neutrinos and
antineutrinos.

In conclusion, we have developed a practical and efficient way of
dealing with lepton number violating neutrino (antineutrino)
oscillations. In a {\it non-adiabatic} perturbation theory, in which
lepton number violation is treated as a small effect compared to
common flavor oscillations, as is compatible with experimental
findings, we have given explicit expressions for the various
oscillation probabilities up to second order in the perturbation and
have embarked upon their interpretation using illustrative examples
for one and two neutrino generations.  For the case of one neutrino
generation, we have found a novel resonance in lepton number violating
oscillations, which in CP non-conserving matter enhances a potentially
small lepton number violation in vacuo.  For the exactly solvable two generation case
of time-independent and flavor blind lepton number violation in vacuo,
we understand that in the presence of lepton number non-conservation
the common flavor oscillation probabilities are periodically modulated
with a frequency given by the lepton number violating coefficient.

\paragraph*{Acknowledgements:} SH is grateful to 
P.B. Pal for inviting him to spend two months at the Saha Institute of
Nuclear Physics, Kolkata, as well as for his warm hospitality during
this period. The bulk of this work was done during this stay in
Kolkata. OM was supported by UEFISCDI project PN-II-RU-TE-2011-3-0184, 
no. 9/05/10/2011.

\end{document}